\documentclass{webofc}
\usepackage[varg]{txfonts}   % Web of Conferences font
\usepackage{amsmath}
\usepackage{rotating}
\usepackage{psfrag}
\usepackage{color}
\usepackage{ifthen}

\pdfoutput=1
\usepackage{graphicx}
\usepackage{dcolumn}
\usepackage{bm}
\usepackage{amsmath}
\usepackage{amssymb}
\usepackage{dsfont}
\usepackage{amsfonts}

\newcommand{\bA}{\boldsymbol{A}}
\newcommand{\bt}{\boldsymbol{t}}

\renewcommand{\vec}[1]{\mbox{\boldmath$#1$\unboldmath}}

\begin{document}
\title{Chiralspin symmetry and its implications for QCD}

\author{\firstname{L. Ya. } \lastname{Glozman}\inst{1}\fnsep\thanks{\email{leonid.glozman@uni-graz.at}} 
        % etc.
}

\institute{Institute of Physics, University of Graz, A-8010 Graz, Austria
          }

\abstract{%
 In a local gauge-invariant theory with massless Dirac fermions a
symmetry of the Lorentz-invariant fermion charge is larger than a symmetry of the  Lagrangian as a whole. While the Dirac Lagrangian exhibits only a chiral symmetry, the fermion charge operator is invariant under a larger symmetry
group, $SU(2N_F)$, that includes  chiral transformations as well as $SU(2)_{CS}$ chiralspin transformations that mix the right- and left-handed
components of fermions. Consequently a symmetry of the electric interaction,
that is driven by the charge density, is larger than a symmetry of the magnetic interaction and of the kinetic term. This allows to separate in some situations
electric and magnetic contributions. In particutar, in QCD the chromo-magnetic
interaction contributes only to the near-zero modes of the Dirac operator,
while confining chromo-electric interaction contributes to all modes. At high
temperatures, above the chiral restoration crossover, QCD exhibits approximate
 $SU(2)_{CS}$ and $SU(2N_F)$ symmetries that are incompatible with free deconfined quarks. Consequently elementary objects in QCD in this regime are
 quarks with a definite chirality bound by the chromo-electric field, without
 the chromo-magnetic effects. In this regime QCD can be described as a stringy fluid. 
}
\maketitle
\section{Introduction}
Here we review recent developments in QCD related to discovery
of a new hidden symmetry in QCD - the chiralspin symmetry \cite{G1,GP}.
The chiralspin symmetry, while not a symmetry of the Dirac
Lagrangian, is a symmetry of the Lorentz-invariant fermion charge.
It allows to separate in some situations electric and magnetic
contributions. This has nontrivial implications on mechanism
of hadron mass generation and on nature of the QCD matter at
high temperatures.

\section{Chiralspin symmetry}
The Dirac Lagrangian with $N_F$ massless flavors

\begin{equation}
{\cal L} = i \bar \psi \gamma_\mu \partial^\mu \psi =
i \bar \psi_L \gamma_\mu \partial^\mu \psi_L +
i \bar \psi_R \gamma_\mu \partial^\mu \psi_R,
\label{DL}
\end{equation}
where 
\begin{equation}
\psi_R = \frac{1}{2}\left( 1+\gamma_5 \right ) \psi,~~
\psi_L = \frac{1}{2}\left( 1-\gamma_5 \right ) \psi,
\label{chirality}
\end{equation}
is chirally symmetric 

\begin{equation}
SU(N_F)_L \times SU(N_F)_R \times U(1)_A  \times U(1)_V.
\label{cs}
\end{equation}
The fermion charge, which is Lorentz-invariant,

\begin{equation}
Q = \int d^3x \bar \psi(x) \gamma_0 \psi(x) = \int d^3x  
\psi^\dagger(x)  \psi(x)
\label{Q}
\end{equation}
is invariant with respect to any unitary transformation
that can be defined in the Dirac spinor space. So
far known unitary transformations were those which leave the Dirac
Lagrangian invariant:

\begin{equation}
SU(N_c),  U(N_F)_L \times U(N_F)_R. 
\label{ls}
\end{equation}

A new unitary transformation of Dirac spinors has recently been found
\cite{G1,GP}. It is a $SU(2)$ chiralspin (CS) transformation.
The  $SU(2)_{CS}$  chiralspin transformation and its generators $\Sigma^n$,
$n=1,2,3$,
are:

\begin{equation}
\psi \rightarrow  \psi^\prime = \exp \left(i  \frac{\varepsilon^n \Sigma^n}{2}\right) \psi  \;,
\label{CS}
\end{equation}

\begin{equation}
 \Sigma^n = \{\gamma_k,-i \gamma_5\gamma_k,\gamma_5\},
\label{SIGCS}
\end{equation}
\noindent 
where $\gamma_k$ is any Hermitian Euclidean gamma-matrix:

\begin{equation}
\gamma_i\gamma_j + \gamma_j \gamma_i =
2\delta^{ij}; \qquad \gamma_5 = \gamma_1\gamma_2\gamma_3\gamma_4.
\label{gamma}
\end{equation}
The $su(2)$ algebra

\begin{equation}
[\Sigma^a,\Sigma^b]=2i\epsilon^{abc}\Sigma^c
\label{algebra}
\end{equation}
is satisfied with any $k=1,2,3,4$.

The $U(1)_A$ chiral symmetry is a subgroup of $SU(2)_{CS}$.
 The free massless Dirac Lagrangian 
is not invariant under the $SU(2)_{CS}$  chiralspin transformation. However, it is a symmetry of the fermion charge.
{\it
The fermion charge  has a larger symmetry than the Dirac equation.}

The chiralspin transformations  and generators 
 can be presented in an  equivalent form. With $k=4$ they are

\begin{equation}
\Sigma^n = \{ \mathds{1} \otimes \sigma^1,~  \mathds{1} \otimes \sigma^2,
 ~  \mathds{1} \otimes \sigma^3 \}.
\label{red}
\end{equation}

\noindent
Here the Pauli matrices $\sigma^i$ act in the  space of spinors

\begin{equation}
 \left(\begin{array}{c}
R\\
L
\end{array} \right),
\label{sp}
\end{equation}

\noindent
where $R$ and $L$  represent the upper and lower components
of the right- and left-handed Dirac bispinors (\ref{chirality}).
The $SU(2)_{CS}, ~k=4$ transformation  can then be rewritten as

\begin{equation}
\label{V-defsp}
  \psi \rightarrow  \psi^\prime = \exp \left(i  \frac{\varepsilon^n \Sigma^n}{2}\right) \psi = \exp \left(i  \frac{\varepsilon^n \sigma^n}{2}\right) \left(\begin{array}{c}
R\\
L
\end{array}\right)\; .
\end{equation}

\noindent
 A fundamental irreducible representation
of $SU(2)_{CS}$ is two-dimensional and the $SU(2)_{CS}$ transformations
mix the $R$ and $L$ components of fermions.

An extension of the direct $SU(2)_{CS} \times SU(N_F)$ product
leads to a $SU(2N_F)$ group. This group contains the chiral
symmetry  $SU(N_F)_L \times SU(N_F)_R \times U(1)_A$ as a subgroup.
Its transformations  are given by

\begin{equation}
\psi \rightarrow  \psi^\prime = \exp\left(i \frac{\epsilon^m T^m}{2}\right) \psi, 
\end{equation}

\noindent
where $m=1,2,...,(2N_F)^2-1$ and the set of $(2N_F)^2-1$ generators is

\begin{align}
T^m=\{
(\tau^a \otimes \mathds{1}_D),
(\mathds{1}_F \otimes \Sigma^n),
(\tau^a \otimes \Sigma^n)
\}
\end{align}
whith $\tau$  being the flavor generators with the flavor index $a$ and $n=1,2,3$ is the $SU(2)_{CS}$ index.
{\it $SU(2N_F)$ is also a symmetry of the fermion charge, while
not a symmetry of the Dirac equation.}

The fundamental
vector of $SU(2N_F)$ at $N_F=2$ is

\begin{equation}
\Psi =\begin{pmatrix} u_{\textsc{R}} \\ u_{\textsc{L}}  \\ d_{\textsc{R}}  \\ d_{\textsc{L}} \end{pmatrix}. 
\end{equation}
\noindent
The $SU(2N_F)$ transformations mix both flavor and chirality.

\section{Symmetries of the QCD action}
Interaction of  quarks with the gluon field in Minkowski space-time
can be splitted into temporal and spatial parts:

\begin{equation}
\overline{\psi}   \gamma^{\mu} D_{\mu} \psi = \overline{\psi}   \gamma^0 D_0  \psi 
  + \overline{\psi}   \gamma^i D_i  \psi ,
\label{cl}
\end{equation}
\noindent
where $D_{\mu}$ is a covariant derivative that includes
interaction of the matter field $\psi$ with the  gauge field $\bA_\mu$,

\begin{equation}
D_{\mu}\psi =( \partial_\mu - ig \frac{\bt \cdot \bA_\mu}{2})\psi.
\end{equation}
The temporal term includes an interaction of the color-octet
 charge density 

\begin{equation}
\bar \psi (x)  \gamma^0  \frac{\bt}{2} \psi(x) = \psi (x)^\dagger  \frac{\bt}{2} \psi(x)
\label{den}
\end{equation}
with the chromo-electric  
part of the gluonic field. {\it
It is invariant  under 
 $SU(2)_{CS}$  and  $SU(2N_F)$} \cite{GP}. 
 The spatial part contains a quark kinetic term
and   interaction with the chromo-magnetic field.  It breaks 
 $SU(2)_{CS}$ and $SU(2N_F)$.  {\it We conclude that  interaction
 of electric and magnetic components
 of the gauge field with fermions can be distinguished
 by symmetry.} Such a distinction does not exist if the matter
 field is bosonic, because a symmetry of the Klein-Gordon Lagrangian
 and of  charge of the $J=0$ field is the same.

Of course, in order to discuss the notions "electric" and "magnetic"
one needs to fix the reference frame. An invariant mass of the hadron
is by definition the rest frame energy. Consequently, to discuss physics
of hadron mass generation it is natural to use the hadron rest frame.

At high temperatures the Lorentz invariance is broken and again
a natural frame to discuss physics is the medium rest frame.

The quark chemical potential term $ \mu \psi (x)^\dagger \psi(x)$ 
in the Euclidean QCD action

\begin{equation}
S = \int_{0}^{\beta} d\tau \int d^3x
\overline{\psi}  [ \gamma_{\mu} D_{\mu} + \mu \gamma_4 + m] \psi,
\end{equation}

\noindent
{\it is $SU(2)_{CS}$  and $SU(2N_F)$
invariant} \cite{G2}.

%%%%%%%%%%%%%%%%%%%%%%%%%%%%%%%%%%%%%%%%%%
\section{Observation of the chiralspin symmetry}

New symmetries presented above were actually reconstructed \cite{G1} from
lattice results on meson spectroscopy upon artificial
subtraction of the near-zero modes of the Dirac operator
from quark propagators \cite{D1}. An initial idea of these
lattice experiments on low-mode truncation was to see whether hadrons survive
or not an artificial restoration of chiral symmetry \cite{LS}.

It is known that the quark condensate of the vacuum,
an order parameter of chiral symmetry, is connected with
the density of the near-zero modes of the Euclidean Dirac
operator via the Banks-Casher relation \cite{BC}

\begin{equation}
 <\bar q q> = -\pi \rho(0).
\end{equation} 
 
\noindent
 The hermitian Euclidean Dirac operator, $i \gamma_\mu D_\mu$, 
 has in a finite volume $V$ a discrete spectrum with real eigenvalues $\lambda_n$:

\begin{equation}
i \gamma_\mu D_\mu  \psi_n(x) = \lambda_n \psi_n(x).
\label{ev}
\end{equation}
Consequently, removing by hands the lowest lying modes
of the Dirac operator from the quark propagators,
 
 \begin{equation}
 S =S_{Full}-
  \sum_{i=1}^{k}\,\frac{1}{\lambda_i}\,|\lambda_i\rangle \langle \lambda_i|,
 \end{equation} 
  
 \noindent
 one artificially restores the chiral symmetry. This truncation of
 the near-zero modes makes the theory nonlocal, but it is
 not a big problem. 
 
\begin{figure}
\centering
\includegraphics[angle=0,width=0.6\linewidth]{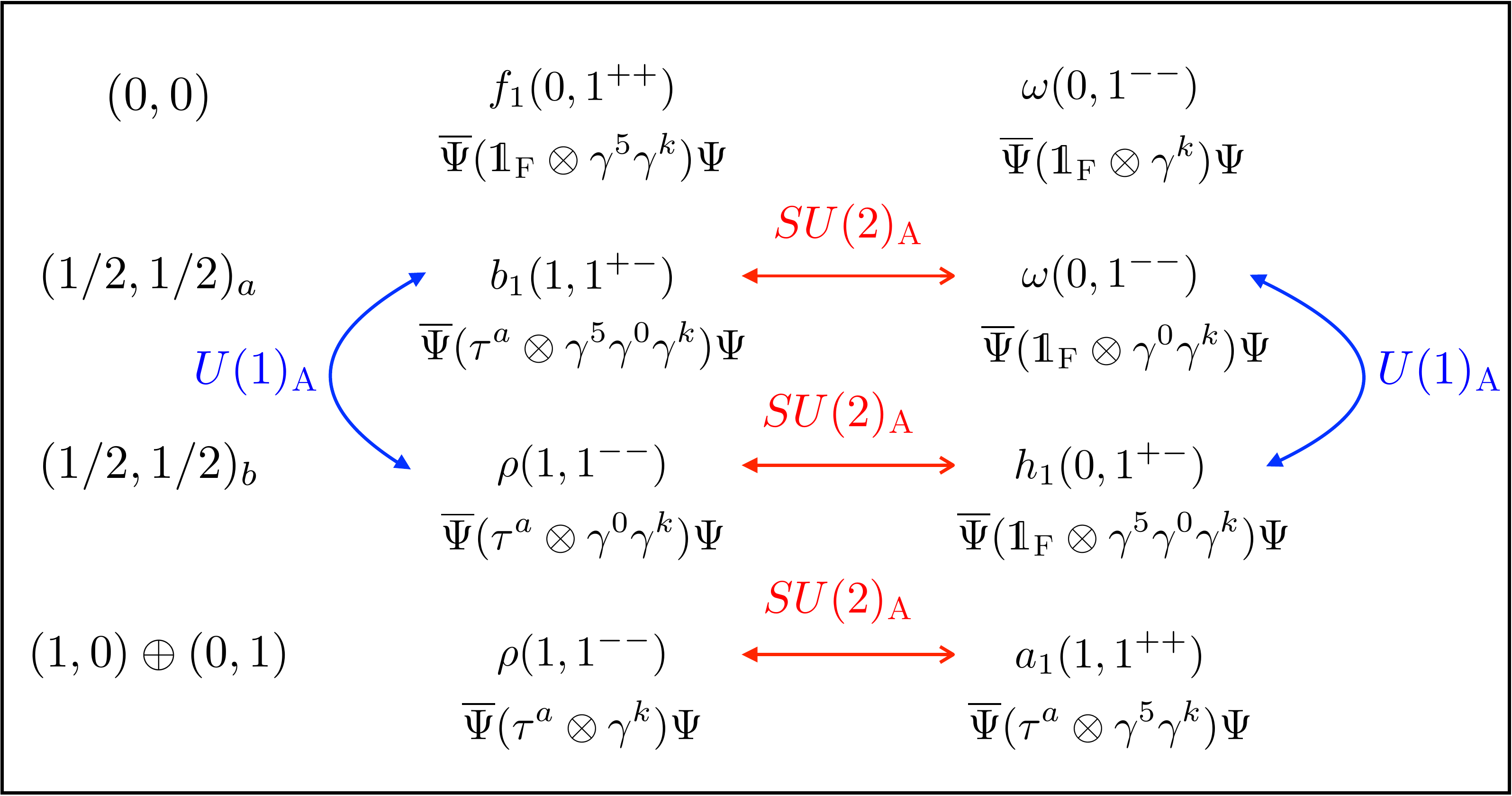}
\caption{$SU(2)_L \times SU(2)_R$ classification of the $J=1$ meson
operators. Operators that are connected by the $SU(2)_L \times SU(2)_R$
and $U(1)_A$ transformations are linked by the red and blue arrows, respectively. The Fig. is from Ref. \cite{GP}.} 
\label{F1}
\end{figure}

 Hadron masses are extracted from the asymptotic slope of the
 rest frame $t$-direction Euclidean correlator
 
 \begin{equation}
C_\Gamma(n_t) = \sum\limits_{n_x, n_y, n_z}
<\mathcal{O}_\Gamma(n_x,n_y,n_z,n_t)
\mathcal{O}_\Gamma(\mathbf{0},0)^\dagger>,
\label{eq:momentumprojection}
\end{equation}
where $\mathcal{O}_\Gamma(n_x,n_y,n_z,n_t)$ is an operator
from the Fig. \ref{F1} that creates a quark-antiquark pair with
fixed quantum numbers.
 Apriori it was not clear whether  hadrons would
 survive the above truncation. If they would, then one should expect a
 clean exponential decay of the Euclidean correlation functions.

A complete set of  $J=1$ local meson operators for $N_F=2$
QCD is given in  Fig. \ref{F1}.
The red arrows connect operators that transform into each other
upon  chiral $SU(2)_L \times SU(2)_R$. If hadrons survive the
artificial chiral restoration then mesons that are connected
by the red arrows should be degenerate. If in addition the $U(1)_A$
symmetry is restored after low mode truncation, then there should
be a degeneracy of mesons connected by the blue arrows. Consequently
from the symmetry of the QCD Lagrangian one can expect degeneracy
of mesons connected by the red and blue arrows. Beyond these arrows
there should be no degeneracy.

\begin{figure}
\centering 
\includegraphics[width=0.55\linewidth]{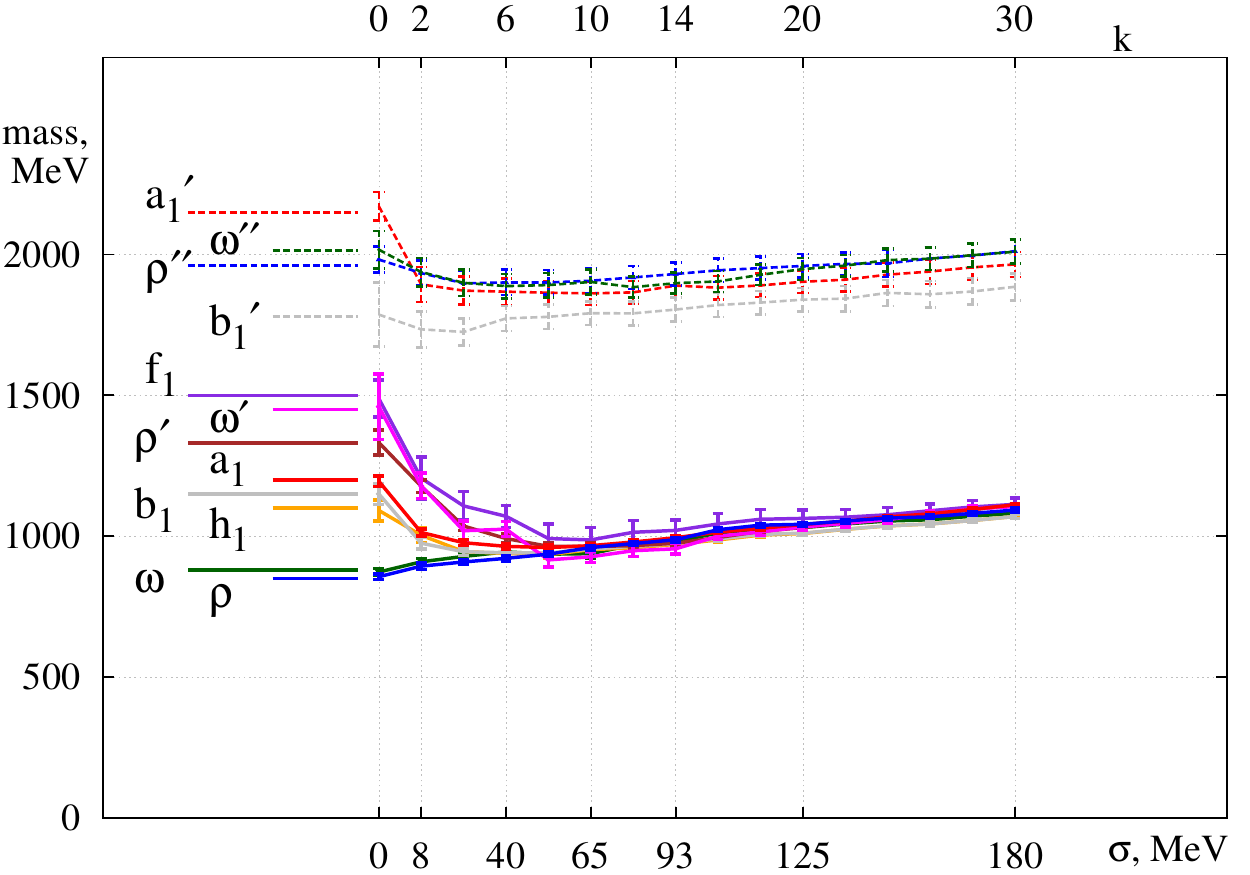}
\caption{$J=1$ meson mass evolution as a function of the
truncation number $k$. $\sigma$ shows the energy gap in the Dirac spectrum. The Fig. is from Ref. \cite{D2}.} 
\label{F2}
\end{figure}

\begin{figure}
\centering
\includegraphics[angle=0,width=0.55\linewidth]{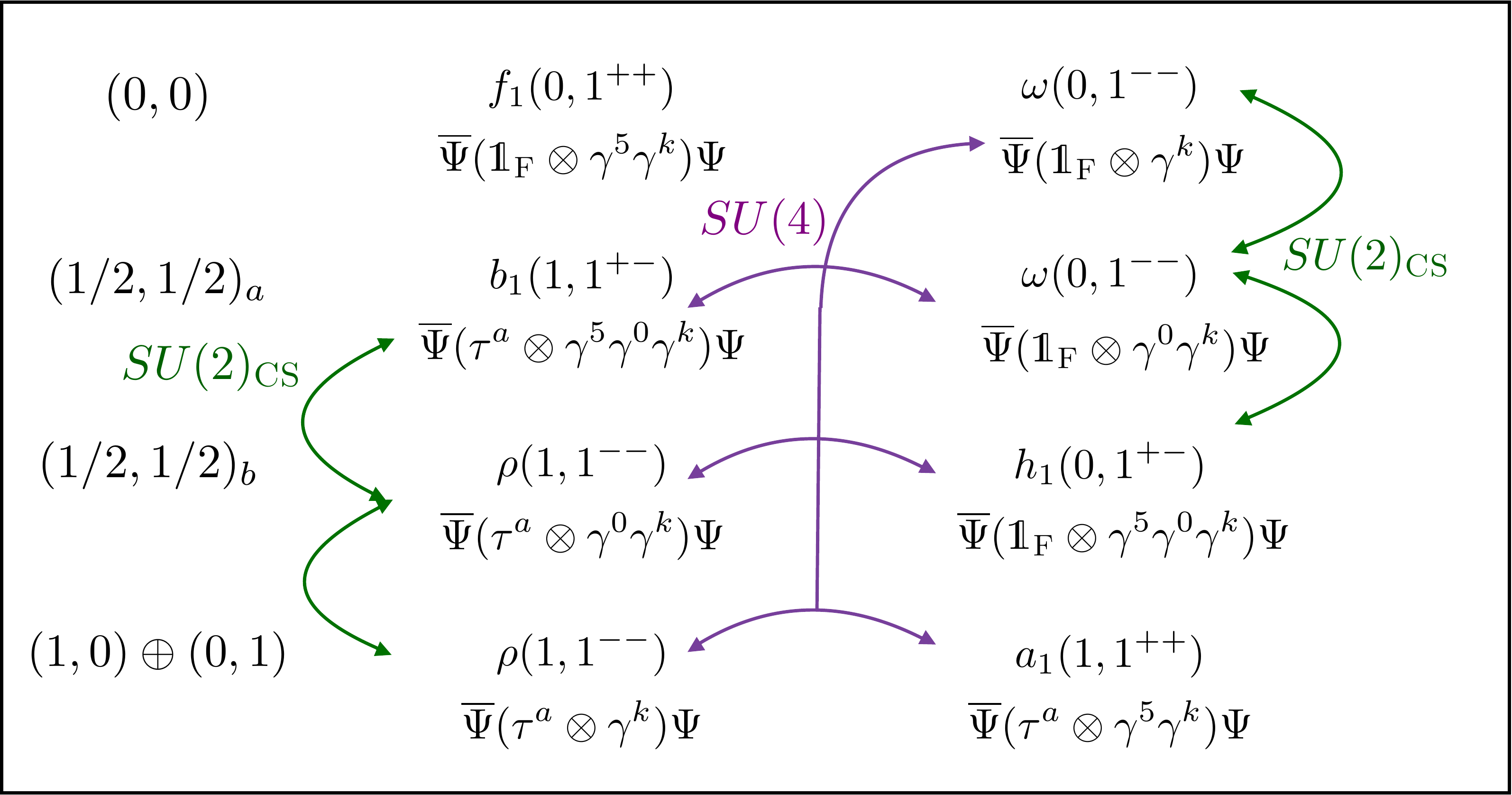}
\caption{The green arrows connect operators that belong to
$SU(2)_{CS}$ triplets. The purple arrow shows the $SU(4)$
15-plet. The $f_1$ operator is is a singlet of $SU(4)$.
The Fig. is from Ref. \cite{GP}.}
\label{F3}
\end{figure}

It was a big surprise when a larger degeneracy than the 
$SU(2)_L \times SU(2)_R \times U(1)_A$ symmetry
of the QCD Lagrangian was found in actual lattice measurements
\cite{D1,D2}, see Fig. \ref{F2}.
This degeneracy, presumably only approximate, represents the $SU(2)_{CS}, ~k=4$  and the $SU(2N_F)$ symmetries because it contains irreducible representations of both groups, see Fig. \ref{F3}.

 The spatial
$O(3)$ invariance is a good symmetry. The $SU(2)_{CS}, ~k=4$
and $SU(2N_F)$ transformations
do not mix bilinear operators from Fig. \ref{F3} with operators
of different spins and thus respect spin of hadrons as a good quantum
number.

These lattice results suggest, given the symmetry classification of
the QCD Lagrangian (\ref{cl}), that while the confining chromo-electric
interaction, that is $SU(2)_{CS}$  and $SU(4)$ symmetric, contributes
 to all modes of the Dirac operator, the
chromo-magnetic interaction, which breaks both  symmetries, is located exclusively in the near-zero
modes. Consequently a truncation of the near-zero modes leads to
emergence of $SU(2)_{CS}$  and $SU(4)$ in hadron spectrum. Similar
results persist for $J=2$ mesons \cite{D3} and baryons \cite{D4}.

The highly degenerate level seen on Fig. \ref{F2} could be considered as a $SU(2N_F)$-
symmetric level of the pure electric confining interaction. The hadron
spectra in nature could be viewed as a  splitting of
the primary level of the dynamical QCD string by means of dynamics
associated with the near zero modes of the Dirac operator, i.e. dynamics
of chiral symmetry breaking that in addition includes all magnetic effects
in QCD.

\section{$SU(2N_F) \times SU(2N_F)$ symmetry of  confining interaction}

 On Fig. \ref{F2} it is well visible that the $SU(4)$ 15-plet and the
 $SU(4)$ singlet ($f_1$) are also degenerate. This means that a confining
 interaction has a larger symmetry that includes the $SU(4)$ as a subgroup.
 What symmetry is it \cite{Cohen}?
 
 The hadron mass is the energy in the rest frame.
 Consider the Minkowski QCD Hamiltonian in Coulomb gauge 
 which is suited to discuss the physics of the hadron mass
 generation in the rest frame \cite{Lee}:

\begin{equation}
H_{QCD} = H_E + H_B
 + \int d^3 x \Psi^\dag({\boldsymbol{x}}) 
[-i \boldsymbol{ \alpha} \cdot \boldsymbol{\nabla} ]  \Psi(\boldsymbol{x})
+ H_T + H_C,
\label{ham}
\end{equation}

\noindent
where the transverse (magnetic) and instantaneous "Coulombic" interactions are:

\begin{equation}
H_T = -g \int d^3 x \, \Psi^\dag({\boldsymbol{x}}) \boldsymbol{\alpha} 
\cdot t^a \boldsymbol{A}^a(\boldsymbol{x}) \, \Psi(\boldsymbol{x}) \; , 
\end{equation}

\begin{equation} 
H_C = \frac{g^2}{2} \int  d^3 x \, d^3 y\, J^{-1} \ \rho^a(\boldsymbol{x})  F^{ab}(\boldsymbol{x},\boldsymbol{y}) \, J \, \rho^b(\bf y) \; ,
\label{coul}
\end{equation}

\noindent
with $J$ being  Faddeev-Popov determinant, $\rho^a(\boldsymbol{x})$ and $\rho^b(\boldsymbol{y})$
are color-charge densities (that include both quark and glue charge
densities) at the space points  $\boldsymbol{x}$
and $\boldsymbol{y}$
and $F^{ab}(\boldsymbol{x},\boldsymbol{y})$ is a 
"Coulombic" kernel. Note that this Hamiltonian  (24)
is not a model but
represents QCD.

While the kinetic and transverse parts of the Hamiltonian are
only chirally symmetric, the confining "Coulombic" part (\ref{coul})
carries the $SU(2N_F)$ symmetry, because the charge density
operator is $SU(2N_F)$ symmetric. However, both $\rho^a(\boldsymbol{x})$ and $\rho^b(\boldsymbol{y})$
 are independently $SU(2N_F)$ symmetric because the $SU(2N_F)$ transformations
 at two different spatial points $\boldsymbol{x}$
and $\boldsymbol{y}$ can be completely independent, with different
rotations angles. Thus the integrand with $\boldsymbol{x} \neq \boldsymbol{y}$
is actually $SU(2N_F) \times SU(2N_F)$-symmetric. 
A contribution with $\boldsymbol{x} =\boldsymbol{y}$
with quarks of the same flavor vanishes because of the Grassmann nature
of quarks.  This means that confining "Coulombic" interaction
 is actually $SU(2N_F) \times SU(2N_F)$-symmetric  \cite{G3}.
 Then it follows that the confining "Coulombic" contributions to all hadrons from an irreducible representation of $SU(2N_F) \times SU(2N_F)$ must
 be the same.

 The $SU(4) \times SU(4)$ has an irreducible representation of dim=16
 that is a direct sum of the 15-plet and of the singlet, $16 = 15 \oplus 1$.
 Then it becomes clear why the 15-plet mesons and the singlet from 
 Fig. \ref{F3} are degenerate in Fig. \ref{F2}.

\section{Topology and the near-zero modes physics}

The physics of the near-zero modes is not only responsible
for chiral symmetry breaking in QCD but also for the breaking
of higher symmetries  $SU(2N_F) \supset SU(2)_{CS}$. The $SU(2)_{CS}$
transformations mix the right- and left-handed components of quarks.
In other words, the physics of the near-zero modes should be
associated not only with breaking of chiral and $U(1)_A$ symmetry,
but also with asymmetry between the left and the right.

Below we suggest a natural microscopic mechanism that induces
an asymmetry between the left- and right-handed components of
quarks in the near-zero modes. This mechanism is related to
the local topological (instanton) fluctuations of the global
gauge configuration.

In a gauge field with a nonzero topological charge the massless fermion
has exact zero modes
\begin{equation}
 \gamma_\mu D_\mu  \psi_0(x) = 0.
\label{dir}
\end{equation}
The zero mode is chiral, $L$ or $R$. 
According to the Atiyah-Singer theorem  the difference of the
number of the left- and right-handed zero modes of the Dirac operator
is related to the topological charge $Q$ of the gauge configuration,

\begin{equation}
Q = n_L - n_R.
\end{equation}

With $|Q| \geq 1$ amount of the right- and left-handed
zero modes is not equal which manifestly  breaks the $SU(2)_{CS}$
symmetry.  The topological configurations
contain the chromo-magnetic field. What would be exact zero modes
in a  topological configuration with a nonzero topological
charge become the near-zero modes of the  Dirac operator in a global
gauge configuration (with arbitrary global topological charge, zero or not zero) that 
contain local topological fluctuations,
like in the Shuryak-Diakonov-Petrov theory of chiral symmetry breaking
in the instanton liquid \cite{S,DP}. Consequently 
on top of contributions from confining physics, which are manifestly  
$SU(2N_F) \supset SU(2)_{CS}$ symmetric, there appear contributions from the topological fluctuations that break $SU(2N_F) \supset SU(2)_{CS}$.

\section{Observation of $SU(2)_{CS}$ and $SU(4)$ symmetries
at high temperatures and their implication}

So far we have discussed $SU(2)_{CS}$ and $SU(4)$ symmetries in hadrons
that emerge upon artificial truncation of the near-zero Dirac modes in $T=0$
calculations. The near-zero modes of the Dirac operator are naturally
suppressed at high temperature above the chiral restoration crossover.  Then, one can expect emergence of $SU(2)_{CS}$ and $SU(4)$ symmetries at high T
without any artificial truncation \cite{G4}.

Given this expectation $z$-direction correlators 

\begin{equation}
C_\Gamma(n_z) = \sum\limits_{n_x, n_y, n_t}
<\mathcal{O}_\Gamma(n_x,n_y,n_z,n_t)
\mathcal{O}_\Gamma(\mathbf{0},0)^\dagger>
\label{eq:momentumprojection}
\end{equation}

\noindent
of all possible $J=0$ and $J=1$
local isovector operators $\mathcal{O}_\Gamma(x) = \bar q(x) \Gamma \frac{\vec{\tau}}{2} q(x)$ have been calculated on the lattice at temperatures 
up to 380 MeV in $N_F=2$ QCD with the chirally symmetric domain wall fermions \cite{R}. The operators and their $SU(2)_L \times SU(2)_R$ and $U(1)_A$ transformation properties are presented in Table~\ref{tab:ops}.
\begin{table}
\center
\begin{tabular}{cccll}
\hline\hline
 Name        &
 Dirac structure &
 Abbreviation    &
 \multicolumn{2}{l}{
   %Symmetries
 } \\\hline
 %%%%%%%%%%%
\textit{Pseudoscalar}        & $\gamma_5$                 & $PS$         & %\multirow{2}{*}
{$\left.\begin{aligned}\\ \end{aligned}\right] U(1)_A$} &\\
\textit{Scalar}              & $\mathds{1}$               & $S$          & &\\\hline
\textit{Axial-vector}        & $\gamma_k\gamma_5$         & $\mathbf{A}$ & %\multirow{2}{*}
{$\left.\begin{aligned}\\ \end{aligned}\right] SU(2)_A$}&\\
\textit{Vector}              & $\gamma_k$                 & $\mathbf{V}$ & & \\
\textit{Tensor-vector}       & $\gamma_k\gamma_3$         & $\mathbf{T}$ & %\multirow{2}{*}
{$\left.\begin{aligned}\\ \end{aligned}\right] U(1)_A$} &\\
\textit{Axial-tensor-vector} & $\gamma_k\gamma_3\gamma_5$ & $\mathbf{X}$ & &\\
\hline\hline
\end{tabular}
\caption{
Bilinear operators and their $SU(2)_L \times SU(2)_R$,  $U(1)_A$
transformation properties. This classification assumes propagation in $z$-direction. The
 index $k$ denotes the components $1,2,4$, \textit{i.e.} $x,y,t$.}
\label{tab:ops}
\end{table}

\begin{figure}
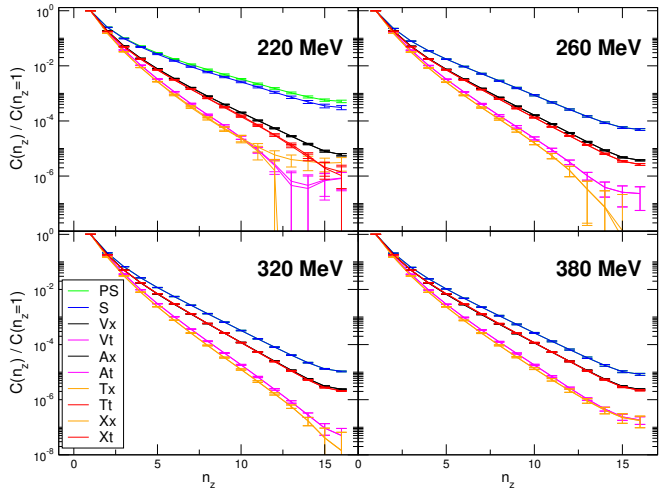

  \centering
  \includegraphics[scale=0.32]{{{1}}}
  \caption{
    Normalized spatial correlators. The Fig. is from Ref. \cite{R}.
  }
  \label{fig:corrs}
\end{figure}

Figure \ref{fig:corrs} shows the  correlators 
for all operators from Table~\ref{tab:ops}.
The argument  $n_z$ is proportional to the dimensionless product
$zT$.  
\noindent
We observe three distinct multiplets:
\begin{eqnarray}
E_1: & \qquad PS \leftrightarrow S \label{eq:e1} \\
E_2: & \qquad V_x \leftrightarrow T_t \leftrightarrow X_t \leftrightarrow A_x \label{eq:e2} \\
E_3: & \qquad V_t \leftrightarrow T_x \leftrightarrow X_x \leftrightarrow A_t. \label{eq:e3}
\end{eqnarray}
$E_1$ is the pseudoscalar-scalar multiplet connected
by the $U(1)_A$ symmetry.
The $E_2$ and $E_3$ multiplets contain however
some operators that are  not connected by 
$SU(2)_L \times SU(2)_R$  or $U(1)_A$
transformations. The symmetries
responsible for emergence of the $E_2$ and $E_3$ multiplets are
$SU(2)_{CS},~k=1,2$ and $SU(4)$, see for details Ref. \cite{R}.
The $SU(2)_{CS},~k=1,2$ transformations do not mix $J=0$ and $J=1$
operators from Table 1 and thus respect spin of mesons
as a good quantum number at high T.

\begin{figure}
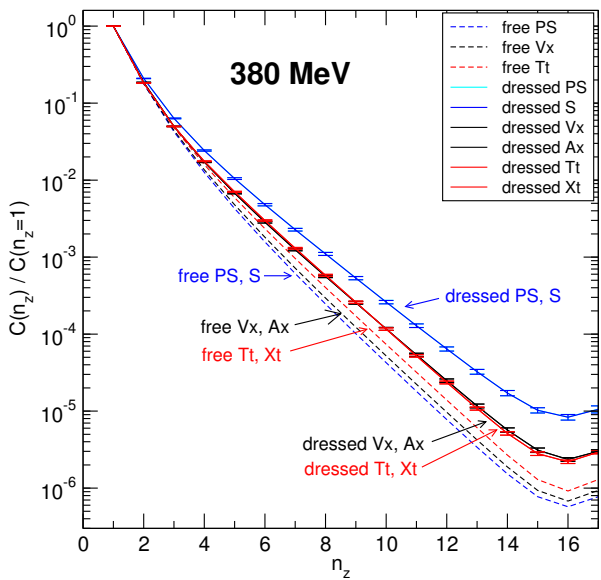

  \centering
  \includegraphics[scale=0.45]{{{2}}}
  \caption{
    $E_1$ and $E_2$ multiplets
    (\protect{\ref{eq:e1}}-\protect{\ref{eq:e2}}) for interacting
    (\textit{dressed}) and non-interacting (\textit{free}) calculations
    at $T$= 380~MeV.
  The Fig. is from Ref. \cite{R}.}
  \label{fig:e2}
\end{figure}

The $U(1)_A$ and $SU(2)_L \times SU(2)_R$ symmetries are exactly or
almost exactly
restored at temperatures above 220 MeV \cite{Cossu:2013uua,Tomiya:2016jwr,
Bazavov:2012qja}. At the same time the $SU(2)_{CS}$ and $SU(4)$ 
 are only approximate. The correlators from the $E_1$ and $E_2$ multiplets
 at the highest 
available temperature 380 MeV are shown in Fig. \ref{fig:e2}.
A remaining $SU(2)_{CS}$ and $SU(4)$ breaking
is at the level of 5\%. 
 We also show there correlators calculated with the noninteracting quarks (abbreviated as "free"). In the latter case only $U(1)_A$ and $SU(2)_L \times SU(2)_R$ symmetries are seen in the correlators.

Scalar (S) and pseudoscalar (PS) systems are bound state systems because
the slopes of the PS and S correlators are substantially smaller than for the free quark-antiquark pair \cite{DeTar:1987xb,Kogut:1998rh}.  In the free
quark case the  minimal slope is determined by twice of the lowest Matsubara frequency 
(because of the antiperiodic boundary conditions for quarks in time direction).
If the quark-antiquark sytem is bound and  of the bosonic nature, the bosonic periodic boundary conditions do allow the slope to be smaller. For the 
$J=1$ correlators the difference of slopes of dressed and free correlators
is smaller  but is still visible. The observed
approximate $SU(2)_{CS}$ and $SU(4)$ symmetries in $J=1$ correlators rule
out asymptotically free deconfined quarks because free quarks do not have
these symmetries. Correlators with such symmetries cannot be obtained
in the weak coupling regime, because perturbation theory relies on a free
Dirac equation that is not $SU(2)_{CS}$- and $SU(4)$-symmetric.

 On Fig. \ref{fig:ratios} we show a ratio of the correlators from the multiplet
$E_2$ 
   at different temperatures. We also show  a ratio calculated with the free noninteracting quarks. For exact $SU(2)_{CS}$ symmetry the ratio should be 1.
At a temperature just above the chiral crossover the ratio is essentially
larger than 1. This can happen only if a contribution from the chromo-magnetic
interaction is still large. Increasing the temperature, the role of the
chromo-magnetic interaction is diminishing and the interaction between quarks
is almost entirely chromo-electric. One concludes that elementary objects
at $T \sim 2T_c$ are not free deconfined quarks, but rather quarks with
a definite chirality bound by the chromo-electric field, something like a string. A remaining small $SU(2)_{CS}$ symmetry breaking is due to
the quark kinetic term in the QCD action.
This conclusion remains also true in matter with finite chemical potentil \cite{G2}.

\begin{figure}
  \centering
  \includegraphics[scale=0.4]{{{4a}}}
  \includegraphics[scale=0.4]{{{4b}}}
  \caption{
    Ratios of normalized correlators ,
    that are related by $U(1)_A$ and $SU(2)_{CS}$ symmetries.
 The Fig. is from Ref. \cite{R}. }
  \label{fig:ratios}
\end{figure}

\begin{figure}
\centering 
\includegraphics[width=0.4\linewidth]{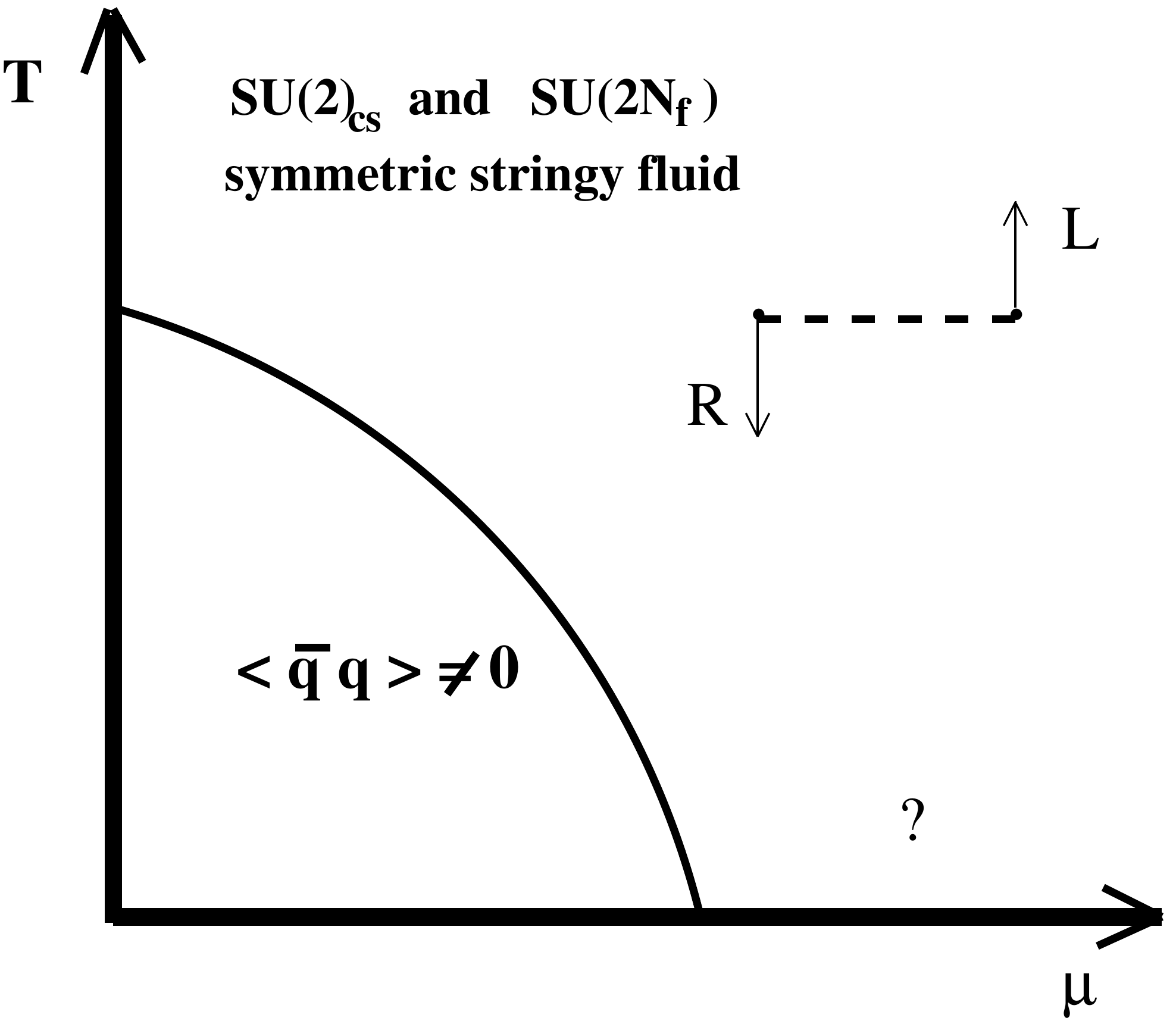}
\caption{A schematic phase diagram}
\label{di}
\end{figure}

How should such a state of matter  be called? It is not
a plasma, because according to the standard definition
plasma is a system of free  charges whith Debye screening of  electric field. From our results it follows
that there are no free deconfined quarks and in addition it is a chromo-magnetic,
but not a chromo-electric field, that is screened. So conditionally one could
call this matter  a stringy fluid, see Fig. \ref{di}. However,  elementary
objects are not usual hadrons. This can be clearly deduced from the correlators
on Fig. \ref{fig:corrs}. E.g. at zero temperature both $V_t$ and $V_x$ operators
couple to one and the same $\rho$-meson. Above the cross-over we
observe that properties of objects that are created  by $V_t$ and $V_x$
operators are very different, because these correlators are very different.
This means that usual $\rho$-mesons get splitted into two independent
objects with not yet known properties. This could explain observed
fluctuations of conserved charges and could  perhaps be experimentally detected
via dileptons.

Preliminary results on correlators at even higher temperatures, up to
1 GeV \cite{Rohrhofer:2018pey}, suggest that at the very high temperatures
the $SU(2)_{CS}$ and $SU(4)$ multiplet structure is washed away and
one approaches to the asymptotic freedom regime.


%=====================================
% References, variant A: internal bibliography
%=====================================
\

% The following MDPI journals use author-date citation: Arts, Econometrics, Economies, Genealogy, Humanities, IJFS, JRFM, Laws, Religions, Risks, Social Sciences. For those journals, please follow the formatting guidelines on http://www.mdpi.com/authors/references
% To cite two works by the same author: \citeauthor{ref-journal-1a} (\citeyear{ref-journal-1a}, \citeyear{ref-journal-1b}). This produces: Whittaker (1967, 1975)
% To cite two works by the same author with specific pages: \citeauthor{ref-journal-3a} (\citeyear{ref-journal-3a}, p. 328; \citeyear{ref-journal-3b}, p.475). This produces: Wong (1999, p. 328; 2000, p. 475)

%%%%%%%%%%%%%%%%%%%%%%%%%%%%%%%%%%%%%%%%%%
\end{document}